\documentclass[bst/sn-mathphys]{sn-jnl}% Math and Physical Sciences Reference Style
\theoremstyle{thmstyleone}%
\theoremstyle{thmstyletwo}%

\theoremstyle{thmstylethree}%

\begin{document}

\title[Anticipated Network Surveillance - An extrapolated study to predict cyber-attacks using Machine Learning and Data Analytics]{Anticipated Network Surveillance - An extrapolated study to predict cyber-attacks using Machine Learning and Data Analytics}


\author*[1]{\fnm{Aviral} \sur{Srivastava}}\email{aviral.srivastava6@s.amity.edu}

\author[2]{\fnm{Dhyan} \sur{Thakkar}}\email{20bec027@nirmauni.ac.in}
\equalcont{These authors contributed equally to this work.}

\author[3]{\pfx{Dr.} \fnm{Sharda} \sur{Valiveti}}\email{sharda.valiveti@nirmauni.ac.in}

\author[3]{\pfx{Dr.} \fnm{Pooja} \sur{Shah}}\email{pooja.shah@nirmauni.ac.in}

\author[3]{\pfx{Dr.} \fnm{Gaurang} \sur{Raval}}\email{gaurang.raval@nirmauni.ac.in}

\affil*[1]{\orgdiv{Department of Computer Science}, \orgname{Amity University}, \orgaddress{ \city{Jaipur}, \postcode{303002}, \state{Rajasthan}, \country{India}}}

\affil[2]{\orgdiv{Department of Electronics and Communications}, \orgname{Nirma Univesity}, \orgaddress{ \city{Ahmedabad}, \postcode{382470}, \state{Gujarat}, \country{India}}}

\affil[3]{\orgdiv{Department of Computer Science}, \orgname{Nirma Univesity}, \orgaddress{ \city{Ahmedabad}, \postcode{382470}, \state{Gujarat}, \country{India}}}


\abstract{Machine learning and data mining techniques are utiized for enhancement of the security of any network. Researchers used machine learning for pattern detection, anomaly detection, dynamic policy setting, etc. The methods allow the program to learn from data and make decisions without human intervention, consuming a huge training period and computation power. This paper discusses a novel technique to predict an upcoming attack in a network based on several data parameters. The dataset is continuous in real-time implementation. The proposed model comprises dataset pre-processing, and training, followed by the testing phase. Based on the results of the testing phase, the best model is selected using which, event class which may lead to an attack is extracted. The event statistics are used for attack prediction. Implementation of the proposed architecture is evaluated against accuracy, precision, recall, and F1 score for processed data.}




\maketitle

\section{Introduction}\label{Intro}

In this age of metaverse, society is witnessing an exponential growth of internet traffic. This leads to the development of indigenous applications, deployment of high-speed networks, and then after, securing of applications, data, and network. As the networks usage increases, sophisticated attacks also increase significantly \cite{regalado2018gray}. These varying attack scenarios give more avenues to the attackers \cite{liao2013intrusion}. The alerts generated by the intrusion detection system are not manageable by the security administrators in real-time \cite{di2008intrusion}. Hence, securing a network in identifying a problem dynamically is very critical.\\
The advances in machine learning and data analysis enabled researchers to automate and extract significant information from the vast data repositories using various sensors (detection systems) in the entire organization \cite{tounsi2018survey}. Hence, the importance of threat intelligence and preparation for attack scenarios become necessary \cite{conti2018cyber}.
\subsection*{Major contributions}
Through this paper, the following outcomes are achieved -
\begin{itemize}
    \item An exhaustive study of various techniques and relevant datasets for developing threat intelligence in a network.
    \item{Proposing a novel threat intelligence architecture model with an intent to anticipate the network attacks}
    \item{Summarizing the performance of different methodologies to implement the said architecture}
    \item{Analyzing the performance of the proposed architecture on various datasets}
    \item{Anticipating the next attack by extracting the key features and analyzing the space of possible events}
\end{itemize}

The paper is organized as follows: Section I\ref{Intro} introduces the proposed concept, Section II \ref{Lit} briefs the work done by researchers in this domain, Section III\ref{Pro} discusses the proposed solution followed by implementation results in Section IV. Conclusions are mentioned in Section V followed by references in Section VI.

\section*{Literature Survey}
\label{Lit}
This section offers gratitude to various researchers for their contribution to the Threat Intelligence domain. In this section, a precise, yet comprehensive survey of various datasets, intrusion predictions, and mathematical, theoretical, and graphical models are explored to emulate the implementation of the proposed threat intelligence system. This survey motivated the design of architecture proposed to anticipate the attacks in the network.

\subsection*{Datasets}
The candidate datasets for the research at hand are various public network logs. They are available for different types of networks with several attacks incorporated within the log. Details of some of the most prevalent data sets are discussed in \cite{valiveti2021anomaly}. Some specific characteristics of these datasets are shown in \ref{tab:datasets}.

\begin{table}[!htp]
    \caption{Summary of datasets with their characteristic details}
    \begin{tabular}{|p{1.7cm}|p{1.6cm}|p{1.1cm}|p{1.8cm}|p{1cm}|p{2.8cm}|}
    \hline
    \textbf{Name of Dataset} & \textbf{No. of features} & \textbf{Year of Creation} & \textbf{Duration of Logs}&\textbf{No. of Classes} & \textbf{Description} \\ \hline
        DARPA 1998 \cite{moustaf2015creating} & 41 & 1998 & 8 weeks & 2 & No representation of real network traffic, false-positive instances missing with irregular in attack data instances \\ \hline
        DARPA 1999 \cite{moustaf2015creating} & 41 & 1999 & 4 weeks & 2 &  "\\ \hline
        DARPA 2000 \cite{moustaf2015creating} & 41 & 2000 &  3 hours, 1:45 hours & 4& "  \\ \hline
        KDD’99 \cite{cup1999data} & 41 & 1999 & Based on Darpa 1999 & 4 & Data samples are redundant and duplicate \\ \hline
        NSL KDD \cite{bala2019review} & 41 & 2000 & Based on KDD'99 & 4 & Another version of KDD CUP 99 dataset with a few attack types\\ \hline
        UNB ISCX 2012 \cite{soheily2018intrusion} & Network Traces & 2012 & 7 Days & 4 & Dataset with network scenarios carrying intrusive activities and labeled data instances  \\ \hline UNSW-NB15 \cite{moustafa2015unsw} & 49 & 2015 & Not available & 10 &  Modern Network Simulated, with Mixed Attack Patterns  \\ \hline
        AFDA \cite{creech2013generation} & System call traces & 2013 & Not available & 73 & Network Intrusion Dataset to represent Modern Networks and Replace KDD \\ \hline
        CICIDS 17 \cite{panigrahi2018detailed} & 80 & 2017 & 5 days & 15 & Customized generation of dataset using Network profiles \\ \hline
        CICIDS 18 \cite{sharafaldin2018detailed} & 80 & 2018 & 10 days & 18 & " \\ \hline
    \end{tabular}
    \label{tab:datasets}
\end{table}

Study concludes that the datasets of DARPA series and KDDCup99 are created using primitive traffic and do not offer better estimates of the modern network traffic. However, UNSW-NB15 and CICIDS offer better diversity of records and mimic the modern real-time network traffic. Thus, they are better suitable for this proposed work.

\subsection*{Threat Intelligence}
Recent innovations in cyber threat intelligence through machine learning techniques are surveyed in /cite{Hus'ak} using use cases of vaticination and soothsaying in the field of cyber security. Researchers worked in the domain of intrusion detection systems (IDS) for various types of networks, offering various applications with customization for organizational requirements. These intrusion detection systems are classified as misuse or pattern-based IDS \cite{kumar1994pattern}, anomaly-based IDS \cite{jyothsna2011review} and specification-based IDS \cite{uppuluri2001experiences}. The efficiency of any IDS is measured in terms of a higher detection rate with reduced false alarms in general \cite{liao2013intrusion}. Alerts are logged and processed. On the contrary, attack anticipation aims to predict an attack based on past studies, extrapolation work, and data analytics \cite{kim2014study} \cite{abdlhamed2017intrusion}. This subsection hence surveys the work leading to attack prediction.

\subsection*{Intrusion Prediction}
”One of the main tasks of intrusion prediction is to detect the novel attacks” \cite{dokas2002data},\cite{lazarevic2003comparative}. Apart from detecting zero-day attacks, intrusion prediction also includes the prediction of vulnerabilities, attack propagation \cite{li2007data} and multi-step attacks \cite{abdlhamed2017intrusion}, and other network security events. While attack protuberance is substantially reckoned on separate models and algorithms of network attacks, there is an abundance of modelsand algorithms which are used for attack prediction. These metods employ different interpretaion of network data which vary from discrete models like attack graphs \cite{jha2002two} or Event Trees to continual approaches like time series methods \cite{abaeian2015intrusion} like ARIMA,GARMA ARMA etc. This makes prediction of attacks and their types possible with the application and assitance of discrete models which are used for attack projection \cite{yang2014attack}. This variation in prediction start aims at considering the odds of exploitation of a particular vulnerability in the network instead of just being an observed malicious event.
\\
The continuous model based time series represents several attacks with reference to network time and request features. The time series will then help to predict the possibility of an attack to happen or not. These continous approaches often include usage of non-technical data along with network logs to predict intrusions through sentiment analysis based methods applicable for social media networks \cite{okutan2017predicting,onoh2018predicting}. Changes in user behaviour can also be supervised to overcome the capriciousness of cyber-attacks.

\subsection*{Forecasting in Network Security}
Situation forecasting is given priority in cyber security to warn the global community against any unpleasant events \cite{husak2018survey} As per Leau et al. \cite{leau2015network}, knowing the complete state of a system or the given network will help us shifting the focus from single attacker in a network. Perception, comprehension, and projection are three levels of network security situation forecasting \cite{shaw2009impact}.
\\
Majority of papers employ quantitative analysis for describing the present state of security \cite{ahmadian2018systematic}. The results obtained through quantitative analysis are used to generate a Network Vulnerability Model in the next phase. No information about precise identification of futuristic attacks is provided by this method. It can, however, issue alerts about a general increase or decrease in securing the network in the future. Quantitative as well as Qualitative analysis necessitates the use of a metric for assessing the security of a network \cite{wang2007measuring}. In the beginning, each host’s security situation is assessed.
\\
Later, the results for each host are factored along with the host’s weight and summed up to get the network’s general security. Authors use different methods for estimating host security. The host’s importance is usually expressed by weight. The quantity and damage incured by attacks on the entire network determine the overall intensity. This prediction can thus serve as a forewarning of an impending change or variation  in the number of attacks as well as their types. The stud
y suggests that since, both the inputs and the predicted values are numerical, almost all the models used to forecast network security situations are continuous models.

\subsection*{Prediction methods in cyber security – A taxonomy}
The methods adopted for attack prediction are enunciated in this subsection. Several approaches \cite{husak2018survey} as mentioned in \ref{fig:taxonomy} are proposed for putting the methods in different categories as shown below:
\begin{figure}
    \centering
    \includegraphics[width =\textwidth]{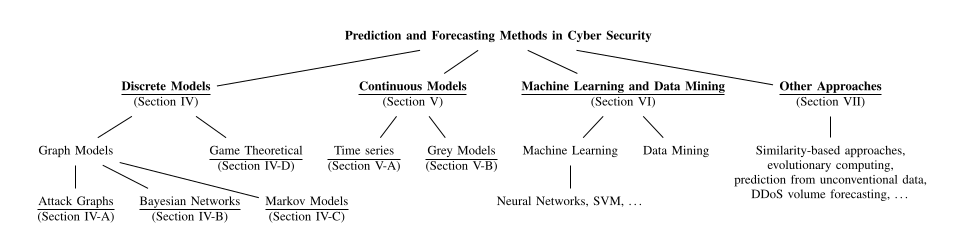}
    \caption{Taxonomy of Methods used for prediction and forecasting of Cyber-Attacks}
    \label{fig:taxonomy}
\end{figure}
\begin{itemize}
    \item Representing an attack or network security situation using theoretical and formal methods.
    \item Methods based on continuous models
    \item Methods employing machine learning and data mining
\end{itemize}
\subsubsection*{Probabilistic models}
This Category of Models explores the attack prediction methods using Probabilistic - graph-based models such as directed attack graphs, Bayesian networks, as well as Markov models.
\subsubsection*{Attack Graph Based}
An attack graph was pioneer method for predicting attacks \cite{li2007data,hughes2003attack}. It is represented as G = (States (S), transition relations (r), Initial State Set (S0), Success state set (Ss)). Transition relations, r, are a delineations of the possible actions which might be used by an attacker with probabilistic weights statistically defined by the action by the attacker. The weights of the paths are then used to predict the past with maximum probability. An attack is counted to be successful if the actions by the attacker lead to any one element of the success state set  from the initial state.
\\
Factor graph is a variant of attack graphs \cite{cao2015preemptive}. A factor graph consists of random variables along with factor functions to determine the weigths.
\subsubsection*{Bayesian Networks}
A Bayesian network is a probabilistic model that represents the variables and the relationships between them \cite{friedman1997bayesian}. The network is a Directed Acyclic Graph (DAG) where the random variables along with their conditional probability weights are enclosed in the nodes.
\\
Let graph G=(Vertex (V), Edges (E)) be a DAG, and where X = (Xi, i ) is a set of random variables indexed by i. Building the model i.e. finding the most appropriate X for the G, requires either expert knowledge about the network and its events, or training using machine learning algorithms. The edge weights are computed using past records. Alert prediction using Bayesian networks uses probabilities represented in the network (model). Probabilities predicted by the network represents the likelihood of the particular event. Low probability alerts are filtered out by appropriate threshold-setting. Only Events with probability higher than the threshold are reported, and a then ppropriate defense mechanisms can be set. Okutanet al. \cite{okutan2017predicting} included signals unrelated to the target network for predicting attacks using the Bayesian network. Accuracy of Prediction ranges from 63\% to 99\%, thereby making the approach quite promising.
\subsubsection*{Markov Models}
Markov models are another popular tool for predicting attacks using model checking prediction methods. Markov models, which include well-known examples of Markov chains \cite{lai2007high,hao2018route} and Hidden Markov Models \cite{haslum2007dips,holgado2017real}, are the popular categories. Being represented using graphs, various algorithms ranging from Bayesian networks to attack graphs can be employed to find insights. Markov models, in contrast to previously published methodologies, work effectively in the presence of any states which are not able to be observed and transition, removing the need for intrusion detection and attack prediction algorithms to have comprehensive information.
\\
In the hidden Markov model (HMM) the state of a model cannot be inferred directly as HMM is modelled as a Markov process with unobserved (hidden) states. In HMM, the outputs depend only on the current state. In case of sequential attacks involving pre-defined events may be detected by an IDS, and therefore the cautions will be raised. With reference to HMMs, these alerts are considered as the observable outputs of attack classes. As IDS cannot descry all events. In order to construct an Hidden Model from the attack sequences, following steps are involved:
\begin{itemize}
    \item Determine cardinatlity of states set, i.e. the total number of states in the model
    \item Determine the number of different observation symbols of each state
    \item Compute the transition probability distribution between the states
    \item Create a distributions of Original states.
\end{itemize}
Observation chances and State transtions are to be extracted manually from records of events \cite{haslum2007dips,holgado2017real}.
\\
Farhadi et al. \cite{farhadi2011alert} proposed a sequential pattern mining approach for generating attack scenarios. These attack scenarios are then used for alert correlation and prediction instead of relying on the manual approach. The attack scenarios are represented using an HMM that is used for attack recognition. This process employs unsupervised machine learning and data mining. Abraham and Nair et al.\cite{abraham2015exploitability} have proposed a predictive framework based on Markov models for exploitability analysis. CVS data has been processed and used to assess the life-cycle of vulnerabilities and predict their impact on the network.

\subsubsection*{Machine Learning and Data Mining Based Models}
The existing methods for applying machine learning techniques for attack prediction are summarized in \ref{tab:mlmethods}.
\begin{table}[]
    \caption{Prevailing Methods for Machine Learning}
    \begin{tabular}{|c|c|}
        \hline
        \textbf{Category} & \textbf{Related Methods}  \\ \hline
         Probabilistic Method & Hidden Markov Model \\ \hline
         Probabilistic Method & Bayesian Network \\ \hline
         Probabilistic Method & Genetic Algorithms \\ \hline
         Intrusion Detection Enhancement & Artificial Neural Networks \\ \hline
         Intrusion Detection Enhancement & Data Mining \\ \hline
         Intrusion Detection Enhancement & Algorithmic Methodologies \\ \hline
         Prediction Methodologies & Alert Correlation with attacks \\ \hline
         Prediction Methodologies & Action / Event Sequence \\ \hline
         Prediction Methodologies & Feature Extraction \\ \hline
    \end{tabular}
    \label{tab:mlmethods}
\end{table}
In most circumstances, machine learning and data mining methodologies are suitable for predicting network security situations. Forecasts of the quantity, volume, and constitution of assaults in the network, as well as their distribution over time, are common outcomes. Here, the work done using time series-based methods is described as continuous data is required to build the prediction model. Cyber-attacks can also be predicted using Spatio-temporal patterns in time series \cite{soldo2011blacklisting}.

\subsubsection*{Time Series Models}
Time series analysis and forecasting is a tool used for predictive analysis which is further used in various applications. Time series are used frequently in Intrusion and Anomaly detection system. A time series is used to represent regular network traffic patterns. As a result, any deviation from the expected value of network parameters (metrics/features) in a given moment are declared as an anomaly.
\\
Historical records of a network phenomenon or statistics, represented numerically is stored as the time series data \cite{husak2018survey}. In the case of cybersecurity and network surveillance, it can be the behavior pattern of an attacker, the requests made by the the attacker or a network security situation state. A lot approaches of time series analysis use moving averages which are calculated by creating a series of averages of subsets of the time series. The weights and exponential smoothing allows the model used to predict to better reflect the nature of the trainng data which is the  input time series.
\\
Abdullah et al. \cite{abaeian2015intrusion} proposed GARMA and Pillai et al. \cite{pillai2015predictive} proposed ARMA time series  was evaluated on live data collected from a honeynet. Zhan \cite{zhan2015predicting} has compared the performance difference between long-term and short-term  predictions of cyber-attacks with various parameters under observation. For this, GARCH and FARIMA time series models were used and about 88\% accuracy was observed with a prediction window i.e. the attack was to take place, after one hour.

\subsubsection*{Machine learning Models}
Methods that use the applications of Machine Learning and Data Mining have found real-world applications allowing the program to learn from data and make decisions without human interference, except to fine-tune hyperparameters for seeking desired outcomes, either in supervised or in unsupervised mode. Long training times and huge computation power are mandated. Machine learning models can be made on a dataset to either regress, cluster, or classify.
\\
Zhang \cite{zhang2012network} illustrated that, at times, a simple network outperforms a complex network as there might be data leaks or a lack of proper data pre-processing steps in place. In \cite{liu2015cloudy} the authors designed a classifier with about 90\% accuracy to prove that machine learning applications can be used for attack detection and may thus be extended to attack forecasting.

\subsubsection*{Tree Based Models}
In the past couple of years, there is a surge in usage of XGboost Models due to their versatility and performance on complex data. \cite{khan2018analysis, disha2022performance, zuech2021detecting, podlodowski2019application}. The GPU Support and performance on XGBoost models is an added advantage for quick and reliable Pipeline design and Development. The prevalence of Random Forests and Boost Models depends on the weak prediction of the estimators which are ensembled. These Estimators are Decision Trees, for limited features and thus allow an ensemble of a large number of them to get such high performance.

\begin{table}[h!]
\caption{Previous Works and their key points}
  \begin{tabular}{|p{2.4cm}|p{1.1cm}|p{2.3cm}|p{1.7cm}|p{3cm}|}
    \hline
    \textbf{Authors}& \textbf{Year}  &\textbf{Approach/Model}   & \textbf{Evalution} & \textbf{Advantages and Outcomes}\\ \hline
    Cao	et	al. \cite{cao2015preemptive} & 2014-2015 & Attack Graph & Live & 75\% Accuracy \\ \hline
    Ramaki et al. \cite{ramaki2015rteca} (RTECA) & 2014 & Attack Graph & DARPA 2000 & 95\% Accuracy \\ \hline
    Okutan	et al. \cite{okutan2017predicting} & 2017 & Bayesian Network & Live & 63-99\%	Accuracy, Non Conventinal Signal \\ \hline
    Farhadi	et al. \cite{farhadi2011alert} & 2011 & Hidden Markov Model & DARPA 2000 &Accuracy of 81.33\% - 98.3\% , Data Mining, Example of a realtime attack Framework \\ \hline
    Abraham and	Nair \cite{abraham2015exploitability} & 2015 & Markov Model & Testbed & Exploitability analysis, vulnerablity life-cycle \\ \hline
    Abdullah , Pillai et al. \cite{abaeian2015intrusion, pillai2015predictive} & 2015 & Time Series (GARMA and ARMA) &Live Data (Honeypot) & Limited Set of Attack Type considered \\ \hline
    Dowling	et al. \cite{dowling2017using} & 2017 & Temporal variances & Live(honeynet) & Attack Type Predictiblity \\ \hline
    Zhang et al. \cite{zhang2012network} & 2013 & BP and RBG neural Network & Custom Dataset & 84.2\%-85.42\% accuracy \\ \hline
    Liu	et	al \cite{liu2015cloudy} & 2015 & Random Forest Classifier & VERIS database, Hackmageddon & Data Breach forecasting \\ \hline
    Khan et al. \cite{khan2018analysis} & 2018 & XGBoost & UNSW-NB15 & 71.7\%	Accuracy and 67	F1-Score \\ \hline
    R. Zuech et al. \cite{zuech2021detecting} & 2021 & Multiple Algorithms & UNSW-NB15 & 93.3 F1 Score \\ \hline
    J.L.Leevy et al \cite{leevy2021detecting} & 2021 & LightGBM and XGBoost & CSE-CICIDS 17 & 95\% Accuracy F1-Score 88.905 \\ \hline
  \end{tabular}
  \label{tab:1}
\end{table}

\subsection*{Challenges}
Following are the challenges in predicting intrusions and mitigating their after effects:
\begin{itemize}
    \item Processing alerts manually is a very slow and error-prone activity
    \item Identifying the point of attack is critical
    \item Mitigating the issue beforehand based on the point of attack is mandated
    \item The larger the network in place, the more complex will be the intrusion prediction and mitigation
    \item Applying contemporary analytics to implement the most suitable learning techniques for real-time intrusion prediction
\end{itemize}
Hence, the proposed architecture addresses the above-mentioned challenges and aims to provide an extrapolated, automatic, real-time intrusion prediction system.
\section*{Proposed Architechure}
\label{Pro}
\begin{figure}
    \centering
    \includegraphics[width =\textwidth]{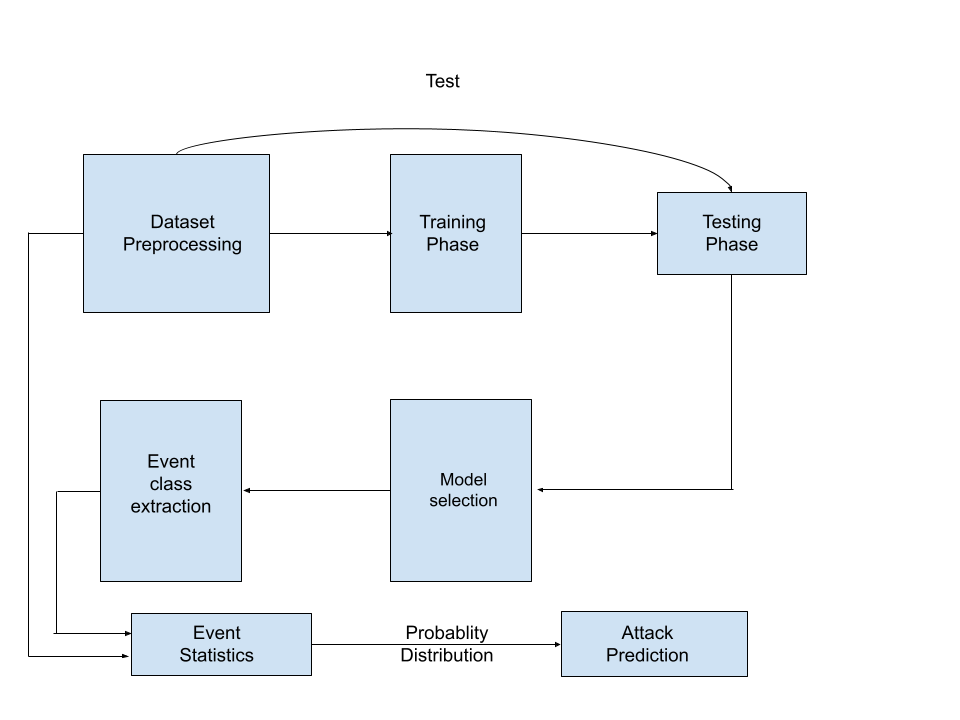}
    \caption{Proposed framework of Intrusion Prediction Process}
    \label{fig:Proposed}
\end{figure}

The proposed architecture solves the mundane task of managing intrusion alerts to identify the point of attack and then hardening the network. The larger the network, the more complex the solution would be. Hence, an approach to automatically predict intrusions with explainable justification about the point of attack that trigger the attack is proposed herewith.
\\
The process is divided into 4 phases as shown in \ref{fig:Proposed}.

\subsection*{Data Preprocessing}
The dataset is analyzed to remove outliers. This dataset, in turn, allows us to have a better understanding and have a more transparent visualization of what the data has to convey. The machine learning models would be able to process the dataset more accurately. In the pre-processing phase, the corrupt entries are removed and new features are created. The new features are derived from the existing features. The focus of the work is on knowing about the attacks instead of the attacker. Hence, the columns related to the source and destination IP are dropped. Dropping the unwanted columns eliminates the horizontals while focusing on the verticals of the dataset. The classifier is hence created for network events which should be able to categorize which event is most likely to occur in the network.

\subsection*{Model Selection}
After the datasets are pre-processed, the model selection phase is initiated. The performance of various algorithms is compared for these datasets. For this task, artificial neural networks, as well as tree-based networks like decision trees and Random Forests, are used. These methods are supplemented by additive boosting and gradient boosting techniques. Probabilistic methods like Naive Bayes and Gaussian process along with the curve fitting methods like simple vector classifiers and logistic regression are used. Accuracy and F1 score are noted to ensure minimum false positives.
\\
Time taken to process the data also happens to be a parameter of concern as the system is supposed to process huge datasets from the modern network infrastructure.
\\
Through the implementation results, the model with a better F1 score and minimum time requirement will be selected to predict the unknown events. The most efficient model will be selected by fine-tuning the hyper-parameters of the model to make it more specific for the network it is being trained for.

\subsection*{Feature Extraction}
\begin{figure}
    \centering
    \includegraphics[width =\textwidth]{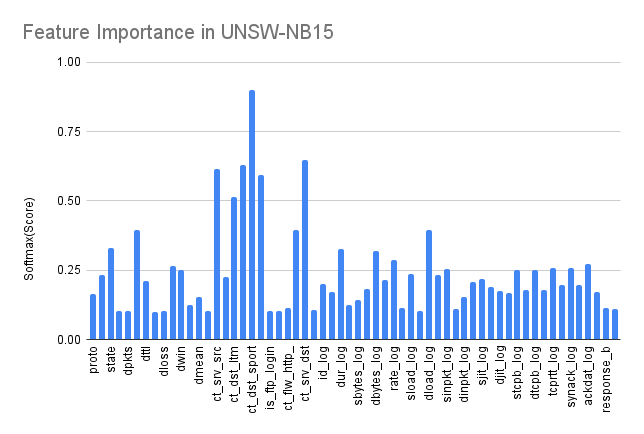}
    \caption{Feature Importance in UNSW-NB15}
    \label{fig:Feature_Importance}
\end{figure}

As of now, a model that almost accurately classifies an event into its respective class is created. These event classes can now be considered separate classes and their features can be extracted. These prime features are nothing but the requirement for any event to most likely be a member of that class. This helps us to create boundaries for the respective classes. For extracting these prime features, Principal Component Analysis along with feature reduction is used. These methods find features that are the most important for correctly classifying the event in their respective classes. Thus, the number of features is reduced with a focus on the significant characteristics of a class which results in further improving the simplicity of the model to allow generalization for unknown events. The importance of features in UNSW-NB15 is shown in \ref{fig:Feature_Importance} and \ref{tab:5}

\subsection*{Hyper-parameter Tuning}
The classifier model with the greatest baseline score is taken for optimization for the concerned dataset to allow better fitting with generalization and specificity of the network model. This re-iterates a pre-made grid of the classifier to allow for better performance on the event space.

\subsection*{Metrics for Model Evaluation}
This subsection explores various performance metrics used to evaluate the developed model.
\begin{figure}
    \centering
    \includegraphics[width =\textwidth]{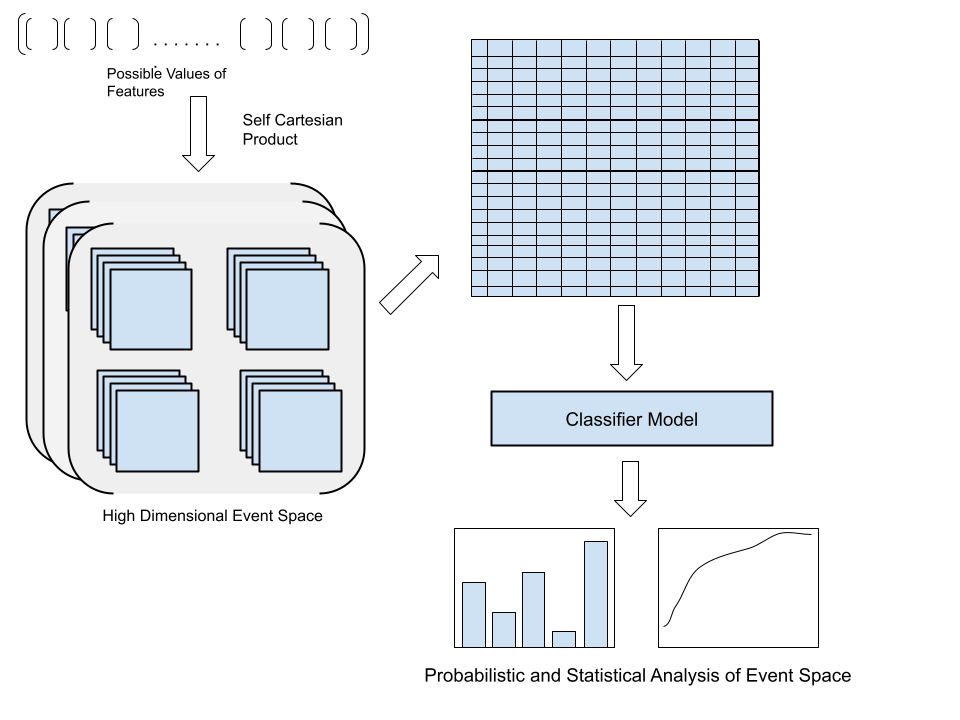}
    \caption{Generation and Analysis of Event Space Results}
    \label{fig:Generation_event_Space}
\end{figure}
\subsubsection*{Attack Prediction}
As the principal features are available for the respective classes, space or domain can be created for all possible events which might happen in the network. Creating this domain as shown in \ref{fig:Generation_event_Space} helps in mapping map the clusters in the higher dimensional space created by the permutations and combinations of the extracted characteristics. This allows the creation of various regions of these events. Using the selected principal components, Cartesian produce is created. This creates a space for all possible events. The space can then be divided into regions of events with a chance to predict the future event type. By finding the probability distribution of these events, we can tell which type/class of attack is the most likely to happen. This probability distribution has been created by extracting prime features from the classes, thereby allowing us to predict what type of attack may happen next and allow us to better prepare for it or at least plan for appropriate mitigation.

\section*{Implementation Results}
\begin{figure}
    \centering
    \includegraphics[width =\textwidth]{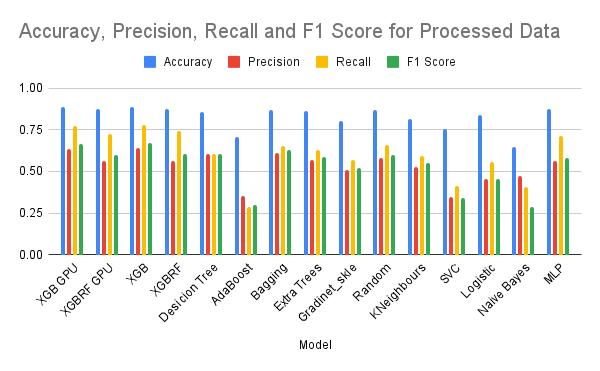}
    \caption{Performance of model on Processed Dataset (UNSW-NB15)}
    \label{fig:Processed_Performance}
\end{figure}

The proposed architecture can predict the type of attack that might happen on the network in the future.
\\
To test the theories, the proposed architecture is implemented on a sample dataset of UNSWNB15 \cite{moustafa2015unsw} and CICIDS-17 \cite{sharafaldin2018detailed}. As the experimentation continues on the given datasets, the scripts are implemented such that they can work independently on any created dataset. As the final classes are categorical, the features are ordinally encoded to a discrete integer mapping with each feature \cite{moustafa2015unsw}.
\\
It is observed that several features of the dataset are continuous with a very large domain. This is not an issue for decision trees; however, it is not appropriate for many other algorithms which work on discrete values. It might be learning just the exact values of the features and making wrong predictions for unseen values, thereby, leading to overfitting. To solve this issue, firstly, new features which can be extracted by the number of zeros in the features or the decimal places in the given feature are plotted into a discrete logarithmic scale. Thus, creating new features and removing the continuous features does allow the model to have a better judgment for significant and minor values of the features while keeping the mapping the same. The continuous signals are mapped into the discrete logarithmic signals and an estimate of the range for a particular entry can be deduced. The change in the number of permutations and combinations is shown in Table 5.
\\
Converting to the logarithmic scale allows us to reduce the number of permutations and combinations of the possible events. This reduces the dimensions of the Cartesian product exponentially. It is because, an entry implies repetition of that value for all permutations and combinations, thereby increasing the dimension of the matrix.
\\
Converting values to the more generalized logarithmic scale instead of the continuous values, not only reduces the total number of combinations, but the algorithm also learns about the range of features. Hence, it introduces more generalization, resulting in better results. The difference can be seen in \ref{tab:2} \ref{fig:Unprocessed_Performance} and \ref{fig:Processed_Performance}:
\begin{figure}
    \centering
    \includegraphics[width =\textwidth]{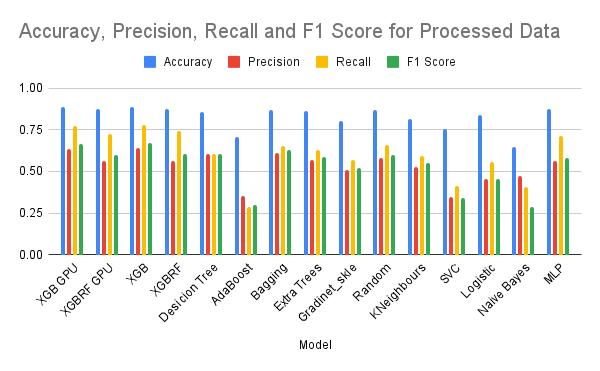}
    \caption{Performance of model on unprocessed Dataset (UNSW-NB15)}
    \label{fig:Unprocessed_Performance}
\end{figure}

To select the most suitable model for this proposed work, various models are deployed and verified against the test data and cross-validation data. Results are evident as shown in \ref{tab:6} and \ref{tab:7}.
\\
Implementation results show that the XGBoost models which are the excessive gradient boosted decision trees perform the best on the training as well as test data. Also, the model consumes less training period which is evident from \ref{tab:3} and \ref{tab:4}.
\\
Next step is the event class extraction as the model for the proposed architecture is selected along with the processed data. As we create the event class extractor, it is realized that the dataset is highly imbalanced. To counter this imbalance, the data is processed again for training the classifier. Various options can be employed to either performing under-sampling or over-sampling or a mixture of both. Under-sampling is a technique to balance uneven datasets by keeping all the data in the minority class and decreasing the size of the majority class (which would be more generalized). Over-sampling means to create artificial entries for the class which has the minimum number of members and thus, creating an even dataset. A combination of under-sampling and over-sampling can fine-tune the classifier without over-fitting it to the data.
\\
After processing the data for the final time and training the event class extractor, the final step is to perform attack prediction.
\subsection*{Attack Prediction}
The accuracy of the model is computed based on the prime / principal features responsible for representing the attack. It can be deduced that, if the model is accurate, this reduced feature-set can be used for modeling the event class predictor.
\\
With all unique features and their individual scores, a Cartesian product can be created to build the model on elements of the given Cartesian product. This would result in a series of attack types. By creating a cumulative distribution of the attack predictions, we have a distribution of the possible attacks.
\\
After receiving the results from the Cartesian product, we can now multiply our accuracy with the results from the distribution, compensating for both the loss at the selected model and the loss which was incurred due to focusing on the prime features of the dataset. Thus, we can get a probability of the future attack class.

\subsection*{Observations}
This section explores the observations and discusses the computation of metrics and their achieved values to justify the implementation results.
\subsubsection*{Evaluation metrics}
\begin{figure}
    \centering
    \includegraphics[width =\textwidth]{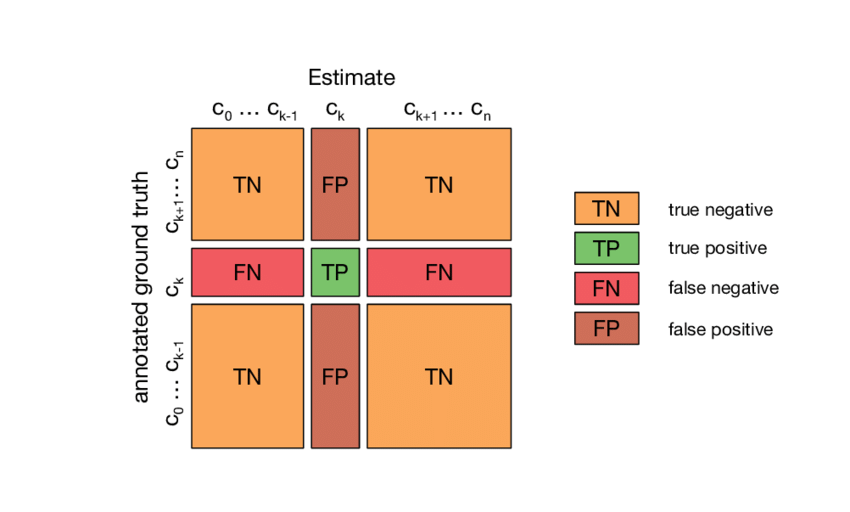}
    \caption{Multi-class Classification Matrix Visualization}
    \label{fig:conf_matrix}
\end{figure}

The multi-class confusion matrix is as shown in \ref{fig:conf_matrix}. This confusion matrix is used to compute the output metrics.

\begin{figure}
    \centering
    \includegraphics[width =\textwidth]{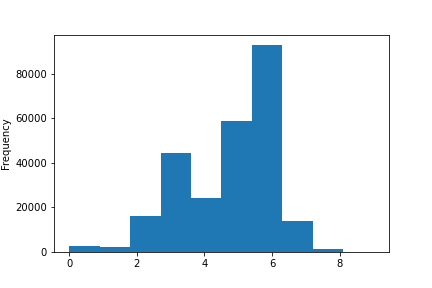}
    \caption{Unbalanced Classes in UNSW-NB15}
    \label{fig:unbalanced_classes}
\end{figure}

\begin{figure}
    \centering
    \includegraphics[width =\textwidth]{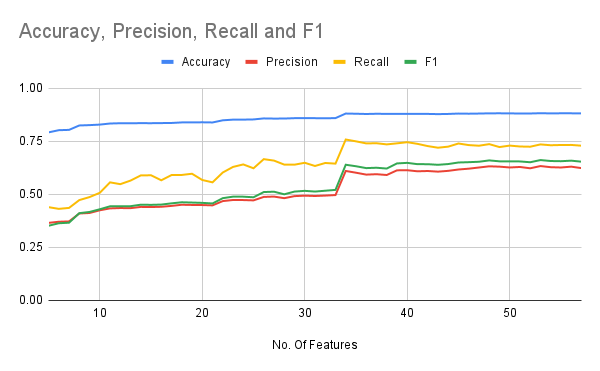}
    \caption{Performance of Model with Feature Reduction (UNSW-NB15)}
    \label{fig:reduction_unsw}
\end{figure}

\begin{figure}
    \centering
    \includegraphics[width =\textwidth]{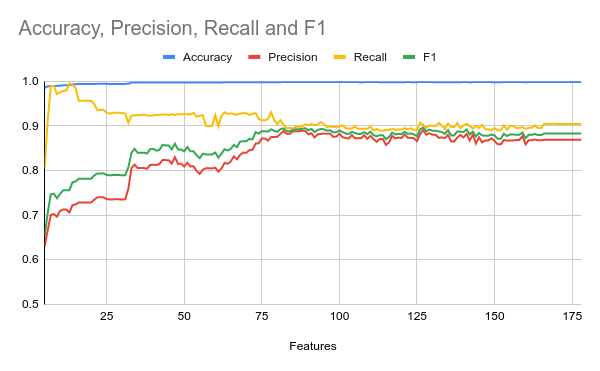}
    \caption{Performance of Model with Feature Reduction (CICIDS-17)}
    \label{fig:reduction_cicds}
\end{figure}

Training and evaluation of model with reference to various metrics, decides how the model learns and thus forms a feedback loop with the model to increase its performance. A combination of accuracy and F1 score are used to check the performance of the system. These metrics are computed as:

\begin{tabular}{l}
 \noalign{\smallskip\smallskip}\hline
 \\
 $Precision = \frac{True\ Positives}{True\ Positives + False\ Positives}$ \\
 \\
 $Recall = \frac{True Positives}{True Positives + False Negatives}$\\
 \\
 $ Accuracy = \frac{True Positives + True Negatives}{All Samples}
 $\\
 \\
 $F1 Score = \frac{2 * Precision * Recall}{Precision + Recall}$
 \\ \\
 \hline
 \end{tabular}
\\
Observing the Probability Distribution of UNSW-NB15 event space as in \ref{fig:unbalanced_classes}, it can be seen that the classes are not uniform. But as we are using Tree based algorithm, specifically XGBoost, the imbalance in the distribution doesn’t affect the performance of the model. This allows for an easier understanding of the network. Detailed statistical analysis of various classes of attacks are mentioned in Additional Files.
\\
\ref{fig:reduction_unsw}and \ref{fig:reduction_cicds} demonstrate the class distribution in event space and change in performance with feature selection in UNSW-NB15 and CICIDS-17.
\section*{Conclusions and Future works}
In this paper, we have presented an architecture for analysing the Event space for a network to assist in predicting the future attack types. The model can also be used for detecting zero day attacks and precisely identifying the fault in the network leading to attack. While the process is quickened by usage of XGBoost algorithm, the statistical analysis of the generated event space is a compute intensive task. Though the computing task is expensive, it allows us to analyse the vulnerability of network without performing experimentation for penetration testing on the existing network. Real-time, circumstantial vulnerability assessment of the network can also be engaged if the suggested model is deployed in a network. Real-time network hardening can also be considered as an upcoming activity under the future work.

\section*{Tables}

\begin{table}[h!]
    \caption{Performance with Variance and Feature Reduction}
    \begin{tabular}{|p{2.3cm}|p{1.4cm}|p{3.4cm}|p{1.6cm}|p{1.6cm}|}
    \hline
    \textbf{Data} & \textbf{Features} &\textbf{ Size Of Event Space (Total Permutations and Combinations)} & \textbf{Accuracy Score} & \textbf{F1 score} \\\hline
    UNSW-Unprocessed & 42 & 9.38e+118 & 0.354 & 0.057 \\ \hline
    UNSW-VarianceReduced & 60 & 5.19e+107 & 0.882 & 0.654 \\ \hline
    UNSW-FeatureReduced & 34 & 4.33e+51 & 0.881 & 0.64 \\ \hline
    CICIDS17-Unprocessed & 78 & 2.40e+287 & 0.999 & 0.813 \\ \hline
    CICIDS17-VarianceReduced & 178 & 1.98e+246 & 0.998 & 0.883 \\ \hline
    CICIDS17-FeatureReduced & 47 & 2.93e+130 & 0.997 & 0.86 \\ \hline
    \end{tabular}
    \label{tab:2}
\end{table}

\begin{table}[h!]
    \caption{Training Period for Processed Data}
    \begin{tabular}{|c|c|}
    \hline
    \textbf{Model} & \textbf{Time}  \\ \hline
    XGB GPU & 4.04 s \\ \hline
    XGBRF GPU & 5.53 s  \\ \hline
    XGB & 6m 32s  \\ \hline
    XGBRF & 6m 13s  \\ \hline
    Decision Tree & 11.1 s  \\ \hline
    AdaBoost & 22.3 s  \\ \hline
    Bagging Classifier & 1m 17s  \\ \hline
    Extra Trees Classifier & 1m 2s  \\ \hline
    Gradient sklearn & 17m 22s  \\ \hline
    Random Forest & 2m 10s  \\ \hline
    KNeighbours & 46m 26s  \\ \hline
    SVC & 31m 42s  \\ \hline
    Logistic Regression & 1h 54m 31s  \\ \hline
    Naive Bayes & 27 ms  \\ \hline
    MLP & 1h 26m 23s   \\ \hline
    \end{tabular}
    \label{tab:3}
\end{table}

\begin{table}[h!]
    \caption{Training Period for Unprocessed Data}
    \begin{tabular}{|c|c|}
    \hline
    \textbf{Model} & \textbf{Time}  \\ \hline
    XGB GPU & 4.9s\\ \hline
    XGBRF GPU & 6.66s  \\ \hline
    XGB & 8m 59s  \\ \hline
    XGBRF & 9m   \\ \hline
    Decision Tree & 4.97 s  \\ \hline
    AdaBoost & 36s  \\ \hline
    Bagging Classifier & 27.4s  \\ \hline
    Extra Trees Classifier & 42.3s  \\ \hline
    Gradient sklearn & 1h17m27s  \\ \hline
    Random Forest & 18m58s \\ \hline
    KNeighbours & 42.3s  \\ \hline
    SVC & 27.1ms \\ \hline
    Logistic Regression & 1h 54m 31s  \\ \hline
    Naive Bayes & 418 ms \\ \hline
    MLP & 1h 21min 4s  \\ \hline
    \end{tabular}
    \label{tab:4}
\end{table}

\begin{table}[h!]
    \caption{Feature Importance of UNSW-NB15}
    \begin{tabular}{|c|c|c|}
    \hline
        \textbf{Feature} & \textbf{K-Best Score} & \textbf{Softmax} \\ \hline
        Proto &5452.93 &0.164 \\ \hline
        Service &11391.53 &0.234 \\ \hline
        State &19550.82 &0.331 \\ \hline
        Spkts &222.4 &0.101 \\ \hline
        Dpkts &421.5 &0.104 \\ \hline
        Sttl &24769.75 &0.394 \\ \hline
        Dttl &9339.65 &0.21 \\ \hline
        Sloss &120.31 &0.1 \\ \hline
        dloss &263.61 &0.102 \\ \hline
        swin &14006.16 &0.265 \\ \hline
        dwin &12911.28 &0.252 \\ \hline
        smean &2149.14 &0.124 \\ \hline
        dmean &4583.7 &0.153 \\ \hline
        trans depth &384.93 &0.103 \\ \hline
        ct srv src &43375.81 &0.615 \\ \hline
        ct state ttl &10692.11 &0.226 \\ \hline
        ct dst ltm &34999.79 &0.515 \\ \hline
        ct src dport ltm &44507.08 &0.629 \\ \hline
        ct dst sport ltm &67280.06 &0.9 \\ \hline
        ct dst src ltm &41603.79 &0.594 \\ \hline
        is ftp login &431.79 &0.104 \\ \hline
        ct ftp cmd &429.49 &0.104 \\ \hline
        ct flw http mthd &1288.6 &0.114 \\ \hline
        ct src ltm &24841.62 &0.394 \\ \hline
        ct srv dst &46007.18 &0.647 \\ \hline
        is sm ips ports &753.38 &0.108 \\ \hline
        id log &8627.27 &0.201 \\ \hline
        id sig &6000.97 &0.17 \\ \hline
        dur log &19177.73 &0.327 \\ \hline
        dur sig &2073.39 &0.123 \\ \hline
        sbytes log &3812.09 &0.144 \\ \hline
        sbytes sig &6964.48 &0.182 \\ \hline
        dbytes log &18444.39 &0.318 \\ \hline
        dbytes sig &9661.28 &0.214 \\ \hline
        rate log &15667.28 &0.285 \\ \hline
        rate sig &1355.59 &0.115 \\ \hline
        sload log &11618.81 &0.237 \\ \hline
        sload sig &211.03 &0.101 \\ \hline
        dload log &24994.61 &0.396 \\ \hline
        dload sig &11170.91 &0.232 \\ \hline
        sinpkt log &13119.91 &0.255 \\ \hline
        sinpkt sig &814.23 &0.108 \\ \hline
        dinpkt log &4665.92 &0.154 \\ \hline
        dinpkt sig &9233.06 &0.209 \\ \hline
        sjit log &9893.81 &0.216 \\ \hline
        sjit sig &7574.17 &0.189 \\ \hline
        djit log &6529.61 &0.176 \\ \hline
        djit sig &5943.85 &0.169 \\ \hline
        stcpb log &12818.59 &0.251 \\ \hline
        stcpb sig &6705.95 &0.178 \\ \hline
        dtcpb log &12807.29 &0.251 \\ \hline
        dtcpb sig &6669.36 &0.178 \\ \hline
        tcprtt log &13363.67 &0.258 \\ \hline
        tcprtt sig &8226.89 &0.197 \\ \hline
        synack log &13368.9 &0.258 \\ \hline
        synack sig &8277.44 &0.197 \\ \hline
        ackdat log &14548.08 &0.272 \\ \hline
        ackdat sig &6205.38 &0.172 \\ \hline
        response body len log &1206.55 &0.113 \\ \hline
        response body len sig &872.83 &0.109 \\  \hline
    \end{tabular}
    \label{tab:5}
\end{table}

\begin{table}[h!]
    \caption{Performance Variation with K-Best Feature Elimination}
    \begin{tabular}{|c|c|c|c|c|}
    \hline
    \textbf{No. Of Features} &\textbf{Accuracy} &\textbf{Precision} &\textbf{Recall} &\textbf{F1} \\  \hline
    5 &0.792 &0.366 &0.44 &0.352 \\ \hline
    6 &0.802 &0.371 &0.432 &0.364 \\ \hline
    7 &0.804 &0.373 &0.436 &0.366 \\ \hline
    8 &0.825 &0.41 &0.473 &0.411 \\ \hline
    9 &0.826 &0.412 &0.487 &0.417 \\ \hline
    10 &0.829 &0.425 &0.507 &0.43 \\ \hline
    11 &0.834 &0.434 &0.557 &0.444 \\ \hline
    12 &0.835 &0.436 &0.548 &0.444 \\ \hline
    13 &0.835 &0.435 &0.565 &0.444 \\ \hline
    14 &0.836 &0.441 &0.589 &0.451 \\ \hline
    15 &0.835 &0.44 &0.59 &0.451 \\ \hline
    16 &0.836 &0.442 &0.566 &0.452 \\ \hline
    17 &0.837 &0.446 &0.591 &0.458 \\ \hline
    18 &0.839 &0.451 &0.592 &0.463 \\ \hline
    19 &0.839 &0.45 &0.597 &0.461 \\ \hline
    20 &0.84 &0.45 &0.568 &0.46 \\ \hline
    21 &0.839 &0.448 &0.557 &0.457 \\ \hline
    22 &0.849 &0.468 &0.603 &0.482 \\ \hline
    23 &0.852 &0.473 &0.629 &0.489 \\ \hline
    24 &0.852 &0.473 &0.641 &0.489 \\ \hline
    25 &0.853 &0.472 &0.623 &0.486 \\ \hline
    26 &0.858 &0.488 &0.666 &0.511 \\ \hline
    27 &0.857 &0.49 &0.659 &0.513 \\ \hline
    28 &0.857 &0.482 &0.64 &0.5 \\ \hline
    29 &0.859 &0.492 &0.64 &0.513 \\ \hline
    30 &0.859 &0.494 &0.649 &0.517 \\ \hline
    31 &0.859 &0.492 &0.634 &0.513 \\ \hline
    32 &0.859 &0.494 &0.648 &0.517 \\ \hline
    33 &0.859 &0.497 &0.645 &0.521 \\ \hline
    34 &0.881 &0.61 &0.759 &0.64 \\ \hline
    35 &0.88 &0.602 &0.75 &0.633 \\ \hline
    36 &0.879 &0.593 &0.74 &0.624 \\ \hline
    37 &0.88 &0.595 &0.741 &0.625 \\ \hline
    38 &0.879 &0.591 &0.736 &0.622 \\ \hline
    39 &0.879 &0.613 &0.741 &0.646 \\ \hline
    40 &0.879 &0.614 &0.746 &0.649 \\ \hline
    41 &0.879 &0.609 &0.738 &0.642 \\ \hline
    42 &0.879 &0.61 &0.728 &0.642 \\ \hline
    43 &0.878 &0.607 &0.72 &0.639 \\ \hline
    44 &0.879 &0.611 &0.725 &0.643 \\ \hline
    45 &0.881 &0.617 &0.74 &0.65 \\ \hline
    46 &0.88 &0.621 &0.732 &0.652 \\ \hline
    47 &0.881 &0.627 &0.729 &0.654 \\ \hline
    48 &0.882 &0.632 &0.737 &0.66 \\ \hline
    49 &0.882 &0.631 &0.723 &0.656 \\ \hline
    50 &0.882 &0.627 &0.73 &0.656 \\ \hline
    51 &0.881 &0.629 &0.726 &0.655 \\ \hline
    52 &0.881 &0.623 &0.725 &0.651 \\ \hline
    53 &0.883 &0.633 &0.735 &0.662 \\ \hline
    54 &0.882 &0.628 &0.732 &0.657 \\ \hline
    55 &0.882 &0.627 &0.733 &0.656 \\ \hline
    56 &0.882 &0.631 &0.733 &0.659 \\ \hline
    57 &0.882 &0.624 &0.729 &0.654 \\ \hline
    \end{tabular}
    \label{tab:6}
\end{table}

\begin{table}[h!]
    \caption{Performance Variation with Recursive Feature Elimination}
    \begin{tabular}{|c|c|c|c|c|}
    \hline
    \textbf{No. Of Features} &\textbf{Accuracy} &\textbf{Precision} &\textbf{Recall} &\textbf{F1} \\  \hline
    5 &0.792 &0.366 &0.44 &0.352 \\ \hline
    6 &0.802 &0.371 &0.432 &0.364 \\ \hline
    7 &0.804 &0.373 &0.436 &0.366 \\ \hline
    8 &0.825 &0.41 &0.473 &0.411 \\ \hline
    9 &0.826 &0.412 &0.487 &0.417 \\ \hline
    10 &0.829 &0.425 &0.507 &0.43 \\ \hline
    11 &0.834 &0.434 &0.557 &0.444 \\ \hline
    12 &0.835 &0.436 &0.548 &0.444 \\ \hline
    13 &0.835 &0.435 &0.565 &0.444 \\ \hline
    14 &0.836 &0.441 &0.589 &0.451 \\ \hline
    15 &0.835 &0.44 &0.59 &0.451 \\ \hline
    16 &0.836 &0.442 &0.566 &0.452 \\ \hline
    17 &0.837 &0.446 &0.591 &0.458 \\ \hline
    18 &0.839 &0.451 &0.592 &0.463 \\ \hline
    19 &0.839 &0.45 &0.597 &0.461 \\ \hline
    20 &0.84 &0.45 &0.568 &0.46 \\ \hline
    21 &0.839 &0.448 &0.557 &0.457 \\ \hline
    22 &0.849 &0.468 &0.603 &0.482 \\ \hline
    23 &0.852 &0.473 &0.629 &0.489 \\ \hline
    24 &0.852 &0.473 &0.641 &0.489 \\ \hline
    25 &0.853 &0.472 &0.623 &0.486 \\ \hline
    26 &0.858 &0.488 &0.666 &0.511 \\ \hline
    27 &0.857 &0.49 &0.659 &0.513 \\ \hline
    28 &0.857 &0.482 &0.64 &0.5 \\ \hline
    29 &0.859 &0.492 &0.64 &0.513 \\ \hline
    30 &0.859 &0.494 &0.649 &0.517 \\ \hline
    31 &0.859 &0.492 &0.634 &0.513 \\ \hline
    32 &0.859 &0.494 &0.648 &0.517 \\ \hline
    33 &0.859 &0.497 &0.645 &0.521 \\ \hline
    34 &0.881 &0.61 &0.759 &0.64 \\ \hline
    35 &0.88 &0.602 &0.75 &0.633 \\ \hline
    36 &0.879 &0.593 &0.74 &0.624 \\ \hline
    37 &0.88 &0.595 &0.741 &0.625 \\ \hline
    38 &0.879 &0.591 &0.736 &0.622 \\ \hline
    39 &0.879 &0.613 &0.741 &0.646 \\ \hline
    40 &0.879 &0.614 &0.746 &0.649 \\ \hline
    41 &0.879 &0.609 &0.738 &0.642 \\ \hline
    42 &0.879 &0.61 &0.728 &0.642 \\ \hline
    43 &0.878 &0.607 &0.72 &0.639 \\ \hline
    44 &0.879 &0.611 &0.725 &0.643 \\ \hline
    45 &0.881 &0.617 &0.74 &0.65 \\ \hline
    46 &0.88 &0.621 &0.732 &0.652 \\ \hline
    47 &0.881 &0.627 &0.729 &0.654 \\ \hline
    48 &0.882 &0.632 &0.737 &0.66 \\ \hline
    49 &0.882 &0.631 &0.723 &0.656 \\ \hline
    50 &0.882 &0.627 &0.73 &0.656 \\ \hline
    51 &0.881 &0.629 &0.726 &0.655 \\ \hline
    52 &0.881 &0.623 &0.725 &0.651 \\ \hline
    53 &0.883 &0.633 &0.735 &0.662 \\ \hline
    54 &0.882 &0.628 &0.732 &0.657 \\ \hline
    55 &0.882 &0.627 &0.733 &0.656 \\ \hline
    56 &0.882 &0.631 &0.733 &0.659 \\ \hline
    57 &0.882 &0.624 &0.729 &0.654 \\ \hline
    \end{tabular}
    \label{tab:7}
\end{table}

\backmatter

\section*{Declarations}

Some journals require declarations to be submitted in a standardised format. Please check the Instructions for Authors of the journal to which you are submitting to see if you need to complete this section. If yes, your manuscript must contain the following sections under the heading `Declarations':

\begin{itemize}
\item Funding- No external Funding was received for the Research Conducted within the Attached work.

\item Competing interests - The authors certify that they have NO affiliations
with or involvement in any organization or entity with any financial interest (such as honoraria;
educational grants; participation in speakers’ bureaus; membership, employment,
consultancies, stock ownership, or other equity interest; and expert testimony or patent-
licensing arrangements), or non-financial interest (such as personal or professional
relationships, affiliations, knowledge or beliefs) in the subject matter or materials discussed in
this manuscript.
\\
The authors declare that they have no competing interests.
\\
\item Ethics approval - Not Applicable
\item Consent to participate - The authors of the manuscript give consent to participate
\item Consent for publication- The authors of the manuscript give consent to publish the manuscript of found suitable
\item Availability of data and materials - The datasets analysed during the current study are available in the at their respective sources whose links are available as follows:\\
DARPA 98 https://www.ll.mit.edu/r-d/datasets/1998-darpa-intrusion-detection-evaluation-dataset \\
DARPA 99 https://www.ll.mit.edu/r-d/datasets/1999-darpa-intrusion-detection-evaluation-dataset \\
DARPA 2000 https://www.ll.mit.edu/r-d/datasets/2000-darpa-intrusion-detection-scenario-specific-datasets\\
KDD’99 http://kdd.ics.uci.edu/databases/kddcup99/kddcup99.html \\
NSL KDD https://www.unb.ca/cic/datasets/nsl.html \\
UNSW-NB15 https://research.unsw.edu.au/projects/unsw-nb15-dataset \\
CICIDS 17 https://www.unb.ca/cic/datasets/ids-2017.html CICIDS 18 \\ https://www.unb.ca/cic/datasets/ids-2018.html \\
\item Code availability - Not Applicable \\
\item Authors' contributions - \\
Aviral Srivastava: Methodology, Data Curation, Correseponding \\
Dhyan Thakkar: Software, Investigation, Writing Original Draft \\
Sharada Valiveti: Formal Analysis, Investigation, Visualization \\
Pooja Shah: Conceptualization, Writing – Reviewing and Editing, Supervision, Project
Administration \\
Gaurang Raval: Writing – Reviewing and Editing, Validation \\
\end{itemize}

\noindent
If any of the sections are not relevant to your manuscript, please include the heading and write `Not applicable' for that section.

\bmhead*{Supplementary information}

If your article has accompanying supplementary file/s please state so here.

Authors reporting data from electrophoretic gels and blots should supply the full unprocessed scans for key as part of their Supplementary information. This may be requested by the editorial team/s if it is missing.

Please refer to Journal-level guidance for any specific requirements.

\bmhead*{Acknowledgments}

Acknowledgements are due to the super-computing infrastructure of Nirma University as we were using the Param Shavak super computer remotely.

\bmhead*{Abbreviations}%% if any
IDS: Intrusion Detection System; ML: Machine learning; RF: Random forest; DT: Decision tree; GBT: Gradient Boosting Tree; FPR: False positive rate; HIDS: Host-based IDS; NIDS: Network-based IDS; SIDS: Signature-based IDS; AIDS: Anomaly-based IDS; DDoS: Distributed Denial of Service; PIO: Pigeon Inspired Optimizer; XGBoost:Extreme Gradient Boosting; LSTM: Long-Short Term Memory; KNN: K-Nearest Neighbors; RBF: Radial Basis Function; SVM: Support Vector Machine; DNN: Deep Neural Network; AUC: Area under the receiver operating characteristic curve; ANN: Artificial Neural Network; PCA: Principal Component Analysis; DoS: Denial of Service; SGD: Stochastic Gradient Descent;


\bibliography{sn-bibliography}% common bib file
\bibliographystyle{IEEEtran}


\end{document}